\documentclass[onecolumn,3p,a4paper,10pt]{elsarticle}
\usepackage{amssymb,natbib,amsmath,wasysym}
\usepackage{newtxtext,newtxmath}
\usepackage{caption}
\usepackage{url}
\usepackage{ulem}
\usepackage{color}
\usepackage{graphicx}
\newcommand{\mnras}{Monthly Notices of the Royal Astronomical Society}
\newcommand{\apj}{The Astrophysical Journal}
\newcommand{\apjs}{The Astrophysical Journal Supplement Series}
\newcommand{\aj}{Astronomical Journal}
\newcommand{\aap}{Astronomy and Astrophysics}

\begin{document}

\begin{frontmatter}

\title{Meshless Methods for Magnetohydrodynamics with Vector Potential}

\author{Xiongbiao Tu$^a$}
\author{Qiao Wang$^{a,b}$}
\ead{qwang@nao.cas.cn}
\author{Haonan Zheng$^{a,b}$}
\author{Liang Gao$^{a,b,c}$}

\address[label1]{National Astronomical Observatories, Chinese Academy of Sciences, Beijing, 100101, China}
\address[label2]{School of Astronomy and Space Science, University of Chinese Academy of Sciences, Beijing, 100048, China}
\address[label3]{
Institute for Computational Cosmology, Department of Physics, Durham University, Science Laboratories, South Road, Durham DH1 3LE, England}

\begin{abstract}
We present a meshless method for magnetohydrodynamics by evolving the vector potential of magnetic fields. A novel scheme and numerical techniques are developed to restrict the divergence of magnetic field, based on the Meshless Finite Mass/Volume with HLLD Riemann solver for conservative flux calculation. We found the magnetic field could be stabilized by a proper smoothing process and so the long-term evolution becomes available. To verify the new scheme, we perform the Brio-Wu shock tube,  2D and 3D Orszag-Tang vortex  and Magnetorotational Instability test problems.  The results suggest that our method is robust and has better precision on central offset, amplitude and detailed pattern than an existing meshless code--GIZMO.
\end{abstract}


\end{frontmatter}

\section{Introduction}
Magnetic fields plays an essential role in various astrophysical processes, from solar physics, accretion disk to galaxy formations~\citep{2018MNRAS.480.5113M, athena2020}. To better match the dynamic range and capture the detail of shock waves and turbulence at small scale, Lagrangian Meshless methods are widely employed in many Astronomical hydro-dynamical simulations. One of the popular schemes, the smoothed-particle hydrodynamics (SPH), was firstly introduced by~\citet{Lucy+1977,Gingold+1977} and has been extensively developed by \citet{Borve2001,Price2004,Springel2005,Cullen2010,Abel2011,Dehnen2012,Hopkins2013}.  Later the Smoothed Particle (or Lagrangian) Magnetohydrodynamics (SPMHD) has been developed  \citep{Maron2003,Price2005,Rosswog2007,Dolag2009,Iwasaki2011,Maron2012,Tricco2012,Stasyszyn2013,Stasyszyn2015} by coupling the gas dynamics with the magnetic field \citep{Hopkins+2016a, Hopkins+2016b, Hawley1995, Toth2000, Miyoshi2005, Stone2008, Li2008, Pakmor2011, Kawai2013, Mocz201403, LOPES2018293}. Recently, the Godunov solver is combined with the meshless methods to improve conservation and avoid some artificial effects of SPH \citep{Lanson+2008a,Lanson+2008b, Gaburov+2011}. For instance, Hopkins developed a new meshless method, GIZMO, which is more effective for the conservation of energy, momenta, and angular momentum\citep{Hopkins+2015, Hopkins+2016a}.

On the other hand, MHD still faces some numerical challenges, for an instance, the divergence-free ($\nabla\cdot\textbf{B}=0$). The conventional conservative schemes are insufficient to keep the divergence of magnetic field being vanished after long-term evolution in MHD simulations.  A Constrained-Transport (CT) method \citep{Evans1988,Fromang2006,Rossmanith+2006,Cunningham2009,Helzel+2011,Miniati2011,Mocz+2014,Mocz+2016,Li2021} is widely adopted in mesh MHD solvers, but it is not available to meshless ones. Correspondingly, \citet{Hopkins+2016b} developed a so-called Constrained-Gradient (CG) scheme for meshless method, which effectively suppress the divergence error, but it still cannot eliminate this problem as effective as CT to mesh method. Moreover, the error of the magnetic field divergence cannot be improved by employing higher resolution, which often causes numerical instability and destroys the correct physical solutions.

In classic electromagnetism, a magnetic field is naturally defined by a vector potential $(\textbf{B} = \nabla \times \textbf{A})$. As an alternative dynamic variable, vector potential $\textbf{A}$ is employed in SPH framework and plasma physics studies \citep{Price+2010, Xiao+2013,Stasyszyn2015, Qin+2016}, instead of $\textbf{B}$. In this work, we introduce a vector potential (VP) method to a meshless MHD framework and combine with Godunov solver in order to achieve a theoretical divergence-free and at the same time better keep conservation.

The structure of this paper is organized as follows. In section 2, we briefly present a detailed theoretical frame of meshless VP method. The technical details and the numerical implementation of the VP method is described in Section 3. Three fundamental numerical experiments are examined in section 4. We draw our conclusion and discuss our results in Section 5.

\section{NUMERICAL METHOD}
\label{sect:2}
\subsection{The magnetohydrodynamic equations}
\label{subsect:1}
In a frame moving with velocity ${\mathbf v}_{\rm f}$, we consider the ideal magnetohydrodynamic equations (going back to pure hydrodynamics in a special condition of $\mathbf{B} = 0$), which describes the physics of an ideal conducting fluid coupling with a magnetic field. Basically, this system composes the continuity, momentum, energy equations, and induction equations, which can be written in the form of hyperbolic conservation laws 
\begin{equation}
  \label{eq:cons}
L_{v}(\textbf{U}) + \nabla\cdot\mathbf{F} = \mathbf{S},
\end{equation}
where $\mathbf{U}$ is the state vector of conserved variables and $L_{v}$ is the transport operator defined as $L_{v}(\textbf{U}) = {\partial \textbf{U}}/{\partial t} + \nabla \cdot( {\mathbf v}_{\rm f}\otimes \textbf{U})$, Conventionally, the notation $\otimes$ is the outer product and $\nabla \cdot \mathbf{F}$ refers to the inner product between the gradient operator and the flux tensor $\mathbf{F} = F  - {\mathbf v}_{\rm f} \otimes\mathbf{U}$. Specifically, their components read
\begin{equation} \label{eq:2.1.2}
\textbf{U}=
\left[\begin{array}{c}
   \rho  \\
   \rho \mathbf{v} \\
   \rho e \\
   \mathbf{B}\\
   \rho \psi
   \end{array}\right], \text{and}~
F =
\left[\begin{array}{c}
   \rho \mathbf{v} \\
   \rho \mathbf{v}\otimes \mathbf{v} + P_{T}\mathcal{I} - \mathbf{B}\otimes\mathbf{B} \\
   (\rho e + P_{T})\mathbf{v} - (\mathbf{v}\cdot \mathbf{B})\mathbf{B} \\
   \mathbf{v}\otimes \mathbf{B} - \mathbf{B}\otimes \mathbf{v}\\
   \rho \psi\mathbf{v}
   \end{array}\right],
\end{equation}
where $\rho$ is density, $u$ is internal energy, $e = u + |\mathbf{B}|^2 /(2\rho) + |\mathbf{v}|^2 /2$ is the total specific energy, 
$P_T = P_{gas} + |\mathbf{B}|^2 /2$ is total pressure including thermal and magnetic pressures, and $\psi$ is an auxiliary scalar field to control the divergence by using the source term $\mathbf{S}$, which can be divided into two terms ~\citep{Powell1999, DEDNER2002645},
\begin{equation} \label{eq:source}
\textbf{S}= \textbf{S}_{\textup{Powell}} + \textbf{S}_{\textup{Dedner}}
=-\nabla\cdot \mathbf{B}\left(\begin{array}{c}
   0  \\
   \mathbf{B} \\
   \mathbf{v}\cdot\mathbf{B}\\
   \mathbf{v}\\
   0
\end{array}\right)
-\left(\begin{array}{c}
   0  \\
   0 \\
   \mathbf{B}\cdot(\nabla\psi)\\
   \nabla \psi\\
   (\nabla\cdot \mathbf{B})\rho c^2_h+\rho\psi/\tau
\end{array}\right),
\end{equation}
where $c_h$ is the characteristic speed and $\tau$ is the damping time scale of the divergence wave (Eq.~10 in \citet{Hopkins+2016a}). 

In this work, we focus on the whether a formulation of MHD coupling to the magnetic vector potential $\mathbf{A}$ can deal with the divergence-free issue. According to Maxwell's equations, the magnetic field is expressed into a curl of potential $A$,
\begin{equation} \label{eq:curl}
\mathbf{B} = \nabla \times \mathbf{A},
\end{equation}
due to the mathematical identity equation $\nabla\cdot(\nabla \times\mathbf{A})\equiv 0$. Thus the evolution of $\mathbf{A}$ is further derived from the induction equation without currents \citep{2000JCoPh.160..649J,2004MNRAS.348..123P}
\begin{equation} \label{eq:2.3.4}
\frac{\partial \mathbf{B}}{\partial t} = \nabla \times (\mathbf{v}\times \mathbf{B}).
\end{equation}
Substituting Eq.~\ref{eq:curl}, we obtain
\begin{equation} 
\frac{\partial \mathbf{A}}{\partial t} = (\mathbf{v}\times \mathbf{B}) + \nabla \phi
\end{equation}
and
\begin{equation} \label{eq:adA}
\frac{d\mathbf{A}}{dt} = (\mathbf{v}\times \mathbf{B}) +  \mathbf{v} \cdot \nabla \mathbf{A}     + \nabla \phi,
\end{equation}
where $\phi$ is an arbitrary scalar function determined by the gauge condition. The component form of Eq.~\ref{eq:adA} is written as follow, the Greek letter denotes the spatial index $\alpha = 1,2,3$, 
\begin{equation} \label{eq:adA2}
\frac{d A^{\alpha}}{dt} = v^1\frac{\partial A^1}{\partial x^{\alpha}} + v^2\frac{\partial A^2}{\partial x^{\alpha}} + v^3\frac{\partial A^3}{\partial x^{\alpha}} + \frac{\partial \phi}{\partial x^{\alpha}}.
\end{equation}
Without loss of generality,  the scalar $\phi$ is fixed to be $0$ (Weyl gauge) in this work. The initial condition of VP is determined by the initial magnetic field up to an additive gradient scalar field. Considering the symmetric or periodic boundaries, we individually choose the integral constant of scalar fields for specific numerical examinations in Sec. 4. In 2-D condition,   the particle motion  and magnetic field $\mathbf{B}$ is constrained at the $x-y$ plane. Thus the components of $A^1, A^2$ and $v^3$ always vanishes and ${ d {A}^{3} }/{d t} \equiv 0$.  Using this extra constraint, we  can  solve the exact evolution of $\mathbf{A}$ and no further numerical error is accumulated into analytical vector potential except for particles shift (see Section~\ref{sect:evp}). 

\subsection{Weak solution \& Conservative Meshless Method}
We consider the weak solution of the conservative equation (Eq.~\ref{eq:cons}) and begin with the definition of inner product of function $f$ and $g$ 
\begin{equation} \notag
<f,g> = \int_{\mathbb{R}^d  \times \mathbb{R}^{+}} fg~ dx dt
\end{equation}
in the range of $x \in \mathbb{R}^d, t \in [0, \infty)$.
To discretize the spatial integral, we employ a continuous symmetric kernel function $W(x-x_{i},h(x))$ constrained by $\int_{\mathbb{R}^d}W(x-x^{\prime},h(x^{\prime}))dx = 1$ at compact support. So that the fraction function $\psi_{i}(x)$ around particle $i$ reads
\begin{equation} \notag
\psi_i(x) = \frac{1}{\omega(x)}W(x-x_i,h(x_i)),
\end{equation}
where $\omega(x) = \sum_{j\in P} W(x-x_j,h(x_j)) $ is the normalization with the kernel size $h(x)$ of function $W$. The  {\bf meshless derivatives} of a scalar function with respect to $x^{\alpha}$, $\alpha \in \{1,2,3\}$ is given by
\begin{equation} 
\begin{split} 
D^{\alpha}_hf(x_i) &= \sum_{j\in P}(f(x_{j})-f(x_i)) \textbf{T}^{\alpha\beta} (x_i)(x_{j}^{\beta}-x_i^{\beta}) \psi_{j}(x_i)\\
&\triangleq\sum_{j\in P}(f(x_{j})-f(x_{i}))\tilde{\psi}_{j}^{\alpha}(x_i),
\end{split}
\label{eq:md}
\end{equation}
where $x_j$ is the position of particle $j$. All particles compose of a discrete particle set $P$ on a compact domain $\Omega$ of $\mathbb{R}^d$.
 The tensor $\textbf{E}^{\alpha\beta}(x_i) = \sum_{j\in P}(x_{j}-x_i)^{\alpha}(x_{j}-x_i)^{\beta}\psi_j(x_i)$ and $\textbf{T}(x_i) = \textbf{E}(x_i)^{-1}$.
Its adjoint operator denoted by a star sign reads
\begin{equation} \notag
D^{\alpha \ast}_hf(x_{i}) = \sum_{j\in P}(f(x_{i})\tilde{\psi}_{j}^{\alpha}(x_{i})-f(x_{j})\tilde{\psi}_{i}^{\alpha}(x_{j})).
\end{equation}

The conservative equation is multiplied by a continuously differentiable test function,
$\varphi \in C_{0}^{1}(\mathbb{R}^d\times\mathbb{R}^{+})$, and integrated over the support
\begin{equation} \notag
\int_{\mathbb{R}^d\times\mathbb{R}^{+}}(\textbf{U}L_{v}^{\ast}(\varphi) + \mathbf{F}(\textbf{U})\cdot\nabla \varphi + \textbf{S}\varphi) dxdt = 0,
\end{equation}
where $L_{v}^{\ast}(\varphi) = {\partial \varphi}/{\partial t} + {\mathbf v}_{\rm f}\cdot\nabla\varphi$. There exists a weak solution to the conservation laws, since
\begin{equation} \notag
<L_{v}(\textbf{U}),\varphi> = - <\textbf{U},L_{v}^{\ast}(\varphi)>.
\end{equation}
We transfer the spatial integral and meshless derivatives of $\varphi$ into the frame of $\mathbf{v}_{\rm f}$ and obtain
\begin{equation}\notag
\int_{\mathbb{R}^{+}} \sum_{i}\left[V_{i}\textbf{U}_{i}\frac{d\varphi_{i}}{dt} + V_{i}\mathbf{F}^{\alpha}(\textbf{U}_{i})D^{\alpha}_h\varphi_{i} + V_{i}\textbf{S}_{i}\varphi_{i} \right] dt = 0.
\end{equation}
Making an integration by parts with respect to $t$ and rearranging the second term of the left hand side, we have the form
\begin{equation} \notag
\int_{\mathbb{R}^{+}} \sum_{i}\left[\frac{d }{dt}(V_{i}\textbf{U}_{i}) + D^{\alpha \ast}_h(V_{i}\mathbf{F}^{\alpha}(\textbf{U}_{i})) - V_{i}\textbf{S}_{i}\right]\varphi_{i} dt = 0.
\end{equation}
This equation holds for any test function $\varphi$, only if the expression inside the parenthesis vanishes, i.e.
\begin{equation} \notag
\frac{d }{dt}(V_{i}\textbf{U}_{i}) + D^{\alpha \ast}_h(V_{i}\mathbf{F}^{\alpha}(\textbf{U}_{i})) = V_{i}\textbf{S}_{i}.
\end{equation}

Considering the pair of particles $i$ and $j$, we calculate the fluxes $\tilde{\mathbf{F}}_{ij}$ on the pair-wise virtual boundaries by a Riemann solver\citep{Gaburov+2011,Hopkins+2015}. The second term is replaced by
\begin{equation}
  \label{eq:2.2.10}
\frac{d }{dt}(V_{i}\textbf{U}_{i}) + \sum_{j\in P}\tilde{\mathbf{F}}_{ij}^{\alpha} \mathbf{\hat{A}}_{ij}^{\alpha} = V_{i}\textbf{S}_{i},
\end{equation}
where $\mathbf{\hat{A}}_{ij}^{\alpha} = V_{i}\tilde{\psi}_{j}^{\alpha} (x_{i})-V_{j} \tilde{\psi}_{i}^{\alpha}(x_{j})$ is the effective area element between particle $i$ and $j$. Meanwhile, the effective volume $V_i$ is approximately equal to ${1}/{\omega(x_i)}$  \citep{Hopkins+2015},
\begin{equation}
  \label{eq:2.2.12}
\begin{split}
 V_i &= \int_{\mathbb{R}^d}\psi_i(x)dx\\
  &= \int_{\mathbb{R}^d}(\frac{1}{\omega(x_i)} + (x-x_i)\cdot\nabla\frac{1}{\omega(x_i)} + \mathcal{O}(h^2(x_i))) W(x-x_i,h(x_i))dx\\
  &= \frac{1}{\omega(x_i)} + \mathcal{O}(h^2(x_i)).
\end{split}
\end{equation}
This expression can be used to compute the spatial integral of an arbitrary function $f(x)$ by a weighted summation over discrete points, 
\begin{equation}
  \label{eq:2.2.13}
\int_{\mathbb{R}^d} f(x)dx = \sum_{i\in P}  f(x_i)\frac{1}{\omega(x_i)} + \mathcal{O}(h^2(x_i)).
\end{equation}

\section{Numerical Implementation}
\subsection{ Meshless discretization}
In this work, we modify the relevant subroutine in GIZMO code to implement our method. Followed \citet{Gaburov+2011,Hopkins2013,Hopkins+2015}, the kernel size $h(x_i)$ for each particle $i$ used in spatial derivative in $\nu$ dimensions, 
\begin{equation}
  \label{eq:NGB}
N_{\textup{NGB}} = C_{\nu}h^{\nu}(x_i)\sum_{j\in P}W(x_j-x_i,h(x_i)),
\end{equation}
where $C_{\nu} = 1, \pi, 4\pi/3$ for $\nu = 1, 2, 3$, and $N_{\textup{NGB}} $ can be set as 20 for the 2-D case, 32 for the 3-D case. The kernel size $h(x_i)$ is calculated iteratively by formulation (\ref{eq:NGB}). Thus, $V_i = \omega^{-1}(x_i) = [\sum_{j\in P}W(x_i-x_j,h(x_j))]^{-1}$. Each particle resides a virtual cell.
The flux through the cell boundaries between two particles is solved by a 1-D Harten-Lax-van Leer Discontinuities (HLLD) Riemann solver. Following the meshless scheme, we reconstruct all those variables at the pair-wise virtual boundaries by their gradients. For instance, at the position around particle $i$, the value of quantity $f$ is linearly reconstructed by $f_{\textup{rec},i}= f_i + (\textbf{x}-\textbf{x}_i)\cdot \nabla f_i$. Because the magnetic field B is still need for the evolution of the other dynamic variables, density $\rho$, pressure $\textup{P}$, velocity $\textbf{v}$. Following \citet{Gaburov+2011,Hopkins+2016a}, the source term $(VS)_i$ in Eq.~\ref{eq:source} for particle $i$ is 
\begin{equation} \notag
(V\mathbf{S})_i=-(V\nabla\cdot \mathbf{B})_i^{\ast}\left(\begin{array}{c}
   0  \\
   \mathbf{B}_i \\
   \mathbf{v}_i\cdot\mathbf{B}_i\\
   \mathbf{v}_i\\
   0
\end{array}\right)
-\left(\begin{array}{c}
   0  \\
   0 \\
   \mathbf{B}_i\cdot(V\nabla\psi)_i^{\ast}\\
   (V\nabla \psi)_i^{\ast}\\
   (V\nabla\cdot \mathbf{B})_i^{\ast}\rho_i c^2_{h,i}+(m\psi)_i/\tau_i
\end{array}\right),
\end{equation}
\begin{equation} \label{eq:divint}
(V\nabla\cdot \mathbf{B})_i^{\ast} = -\sum_j\bar{\mathbf B}_{x,ij}^{\prime}|\mathbf{\hat{A}}_{ij}|,
\end{equation}
\begin{equation} \notag
(V\nabla\psi)_i^{\ast} = -\sum_j\bar{\psi}_{ij}\mathbf{\hat{A}}_{ij},
\end{equation}
where $\bar{\mathbf B}_{x,ij}^{\prime}$ and $\bar{\psi}_{ij}$ are the combination of the normal component of  the magnetic field ${\mathbf B}_{x}^{\prime}$ with respect to area elements and $\psi$, their forms read
\begin{equation} 
\bar{\mathbf B}_{x,ij}^{\prime} = \frac{1}{2}({\mathbf B}_{x,L}^{\prime} + {\mathbf B}_{x,R}^{\prime}) + \frac{1}{2\tilde{c}_{h,ij}}(\psi_L - \psi_R),
\end{equation}
\begin{equation} 
\bar{\psi}_{ij} = \frac{1}{2}(\psi_L + \psi_R) + \frac{\tilde{c}_{h,ij}}{2}({\mathbf B}_{x,L}^{\prime} - {\mathbf B}_{x,R}^{\prime}).
\end{equation}
We set ${\mathbf{v}}_{x}^{\prime}$ as  projected speed with respect to the frame, then $c_h = {\mathbf{v}_{x}^{\prime}} + c_+$, ${\mathbf{v}_{x}^{\prime}} + c_\pm$ is the fast/slow magnetosonic wave speed, 
\begin{equation} \label{eq:2.1.4}
c_\pm=  \left[\frac{a^2}{2} + \frac{|\mathbf B|^2}{2\rho} \pm \sqrt{\left(\frac{a^2}{2}  +\frac{|\mathbf B|^2}{2\rho} \right)^2 - \frac{a^2{\mathbf B}_{x}^{\prime 2}}{\rho}} \right]^{\frac{1}{2}}.
\end{equation}
Here, $a=\sqrt{{\gamma P}/{\rho}}$ is the sound speed, $\tilde{c}_{h,ij} = \max\{c_{+,i},c_{+,j}\}$, and $\tau_i = L_i/(c_rc_{h,i})$, $L_i$ is an effective size of the particle ($L_i = V_i,\sqrt{2V_i/\pi},(3V_i/4\pi)^{1/3}$ for 1D, 2D, and 3D, respectively) and the constant $c_r$ is equal to 0.03 ~\citep{MIGNONE20102117}. We take $c_{h,i}=v_{sig,i}^{\rm Max}/2$, where $v_{sig,i}^{\rm Max}$ is the maximum signal velocity \citep{Tricco2012,Hopkins+2016a}.  

According to the definition of meshless derivative (Eq.~\ref{eq:md}), the discrete divergence condition is numerically satisfied. Therefore, the instead evolution of VP would not cause a long term bias. Although the numerical differential form of the divergence of the magnetic $\nabla_h \cdot\mathbf B = 0$, the integrated magnetic flux is not zero, especially across the magnetic shock. So the divergence-cleaning term  (Eq.~\ref{eq:divint}) is inevitable to control numerical stability in VP method.

\subsection{{Evolution of vector potential}}

\label{sect:evp} 
In the original GIZMO method, the magnetic field is advanced by using HLLD flux to solve the equation of magnetic field $\mathbf{B}$ in MHD equation (\ref{eq:cons}). But in our method, magnetic field is derived from vector potential at every single step.

In the three dimensional MHD, the magnetic field $\mathbf{B}$ is directly determined by Eq.~\ref{eq:curl}. According to Eq.~\ref{eq:adA2}, the explicit scheme for the evolution of vector potential $\mathbf{A}$ reads
\begin{equation}
\mathbf{A}_i^{n+1} = \mathbf{A}_i^{n} + \Delta t ([\nabla\mathbf{A}_i^{n}]^{\top}\mathbf{v}_i^{n+1/2}), 
\label{eq:ad1}
\end{equation}
where $\mathbf{v}_i^{n+1/2}$ is the median value of the velocity of leapfrog scheme. 

Unfortunately, using simple scheme of Eq.~\ref{eq:ad1}, the long-term evolution of magnetic field is numerically unstable.
In the three dimensional MHD, the evolution of vector potential $\mathbf{A}$ is not analytical, the error of  magnetic field $\mathbf{B}$ derived by Eq.~(\ref{eq:curl}) ruins the stability of numerical solution. To overcome this problem, we introduce a convolution of the magnetic $\mathbf{B}$. As follow,
the smooth function $f_{W}(x)$ of the function $f$ can be written as
\begin{equation} \label{eq:Regular}
f_{W}(x) = \int_{\mathbb{R}^n}f(x-y)W(y)dy = \int_{\mathbb{R}^n}f(y)W(x-y)dy.
\end{equation}
Here, $W(x)$ is a  kernel function. The function $f_{W}(x)$ converges to the function $f$ as the kernel size decreasing. We refer to magnetic field by the Eq.~\ref{eq:curl} as   {$\mathbf{B}_{\rm d}$}, then calculate the regularizing magnetic $\mathbf B$ from the discrete magnetic field {$\mathbf{B}_{\rm d}$} by Eq.~\ref{eq:Regular}, i.e.,
\begin{equation} \label{eq:finalB}
\mathbf{B}(x) = \int_{\mathbb{R}^n}\mathbf{B}_{\rm d}(y)W(x-y)dy, ~ \mathbf{B}_{\rm d}(x) = \nabla\times \mathbf{A}(x).
\end{equation}
which is directly calculated by the sum of $\sum_{j\in P}V_i\mathbf{B}_{\rm d}(x_j)W(x_i-x_j,h(x_i))$. It is easy to tell that the numerical convergence is not changed. Hence, the evolution of VP is numerically governed by the discrete scheme (of  Eq.~\ref{eq:adA} ) as 
\begin{equation}
\label{eq:ad_3d_A}
\mathbf{A}_i^{n+1} = \mathbf{A}_i^{n} + \Delta t (\mathbf{v}_i^{n+1/2}\times \mathbf{B}_i^{n} + \mathbf{v}_i^{n+1/2} \cdot \nabla\mathbf{A}_i^{n}   ). 
\end{equation}
Compared with Eq.~\ref{eq:ad1}, it involves the smoothed magnetic field. Such a stabilizing procedure could avoid the numerical oscillation of $\mathbf{B}$ field, during the evolution. This smoothing process for the discrete magnetic field corresponds to the similar method in \citet{Stasyszyn2015}.

In 2-D case, there is an extra simplification to Eq.~\ref{eq:ad_3d_A}. Since we already know that $A^1=0$, $A^2=0$ and ${d {A^3}}/{d t} \equiv 0$ in the end of section~\ref{subsect:1}. The vector potential of particle $i$ is bound with particle motion,
\begin{equation}
\mathbf{A}_i^{n+1} = \mathbf{A}_i^{n}.
\end{equation}
It means that $\mathbf{A}$ evolves analytically and no numerical error is introduced by vector potential in the numerical modelling, compared with the magnetic field. It is the simplest Lagrangian scheme and a perfect solver to conserve the good behavior of dynamical variables. We use this scheme for 2D Orszag-Tang vortex examination in section~\ref{sec:2dotv}.

\subsection{Time stepping} 
We use the leapfrog scheme for temporal advance with so called Kick-Drift-Kick (KDK) algorithm:
\begin{equation}
  \label{eq:discrete1}
\text{Kick}:\left\{
\begin{aligned}
m_i^{n+1/2} &= m_i^{n} + \dfrac{\Delta t}{2}\left[\dfrac{dm_{i}}{dt}\right]^{n}\\
\mathbf{v}_i^{n+1/2} &= \mathbf{v}_i^{n} + \dfrac{\Delta t}{2}\left[\dfrac{d\mathbf{v}_{i}}{dt}\right]^{n}\\
e_i^{n+1/2} &= e_i^{n} + \dfrac{\Delta t}{2}\left[\dfrac{de_{i}}{dt}\right]^{n},
\end{aligned}
\right.
\end{equation}

\begin{equation}
  \label{eq:discrete2}
  \text{Drift}:\left\{
  \begin{aligned}
\mathbf{x}_i^{n+1} &= \mathbf{x}_i^{n} + \Delta t \mathbf{v}^{n+1/2}\\
\mathbf{A}_i^{n+1} &= \mathbf{A}_i^{n} + \Delta t (\mathbf{v}_i^{n+1/2}\times \mathbf{B}_i^{n} + \mathbf{v}_i^{n+1/2} \cdot \nabla\mathbf{A}_i^{n}   ),
\end{aligned}
\right.
\end{equation}

\begin{equation}
  \label{eq:discrete3}
\text{Kick}:\left\{\begin{aligned}
m_i^{n+1} &= m_i^{n+1/2} + \dfrac{\Delta t}{2}\left[\dfrac{dm_{i}}{dt}\right]^{n+1}\\
\mathbf{v}_i^{n+1} &= \mathbf{v}_i^{n+1/2} + \dfrac{\Delta t}{2}\left[\dfrac{d\mathbf{v}_{i}}{dt}\right]^{n+1}\\
e_i^{n+1} &= e_i^{n+1/2} + \dfrac{\Delta t}{2}\left[\dfrac{de_{i}}{dt}\right]^{n+1},
\end{aligned}\right.
\end{equation}
where $\left[\dfrac{dm_{i}}{dt}\right]^{n+1}$, $\left[\dfrac{d\mathbf{v}_{i}}{dt}\right]^{n+1}$ and $\left[\dfrac{de_{i}}{dt}\right]^{n+1}$ are computed by discrete conservative equation (Eq. \ref{eq:2.2.10}), as follow
\begin{equation}
  \label{eq:pred}
\left[\frac{d(V_{i}\textbf{U}_{i})}{dt}\right]^{n+1} = \left[V_{i}\textbf{S}_{i} - \sum_{j\in P}\tilde{\mathbf{F}}_{ij}^{\alpha} \mathbf{\hat{A}}_{ij}^{\alpha} \right]^{pred}.
\end{equation}
The quantities marked with the superscript "$pred$" depends on the mass $ m_i^{pred} = m_i^{n} + {\Delta t}\left[{dm_{i}}/{dt}\right]^{n}$, velocity $ \mathbf{v}_i^{pred} = \mathbf{v}_i^{n} + {\Delta t}\left[{d\mathbf{v}_{i}}/{dt}\right]^{n}$ and energy $ e_i^{pred} = e_i^{n} + {\Delta t}\left[{de_{i}}/{dt}\right]^{n}$. The flux $\tilde{\mathbf{F}}_{ij}^{\alpha}$ is solved by HLLD Riemann solver. The regularizing magnetic $\mathbf{B}$ estimated by $\mathbf{A}_i^{n+1}$ enters into Eq. \ref{eq:pred}. All numerical gradient estimators use the formula (\ref{eq:md}) in this work.

\section{NUMERICAL TESTS}
\label{sect:analysis}
In this section, we run a series of 2-D and 3-D test problems to examine the accuracy our method,  and compare with some existing works, e.g. GIZMO with the constraint gradient (CG) correction, in Meshless Finite Mass (MFM) and Volume (MFV) methods, respectively. All methods GIZMO, CG, VP-MFM and VP-MFV use the source terms of $\textbf{S}_{\textup{Powell}}$ and $\textbf{S}_{\textup{Dedner}}$ in formula (\ref{eq:source}). The $\textbf{S}_{\textup{Powell}}$ contributes to the numerical stability and $\textbf{S}_{\textup{Dedner}}$  transports the error of numerical magnetic flux away and damps it.

\subsection{Brio-Wu shocktube}
We first perform the Brio-Wu shocktube test. In the 2D periodic boundary condition (box size of $x \in (0,4)$ and $y \in(0,0.25)$), we employ $896\times56$ particles and setup left-state vector as $(\rho, v^1, v^2, v^3, B^1, B^2, B^3, P) = \{1, 0, 0, 0, 0.75, 1, 0, 1\}$  and right-state $\{0.125, 0, 0, 0, 0.75, -1, 0, 0.1\}$, following the initial condition in \citet{Brio+1988}. The magnetic is determined by the vector potential of $A^3 = 0.75y \mp (x-2)$ for left and right parts, respectively. The index of equation of state is $\gamma =2$.
\begin{figure*}[htpb] 
\centering
\includegraphics[width=1.\linewidth]{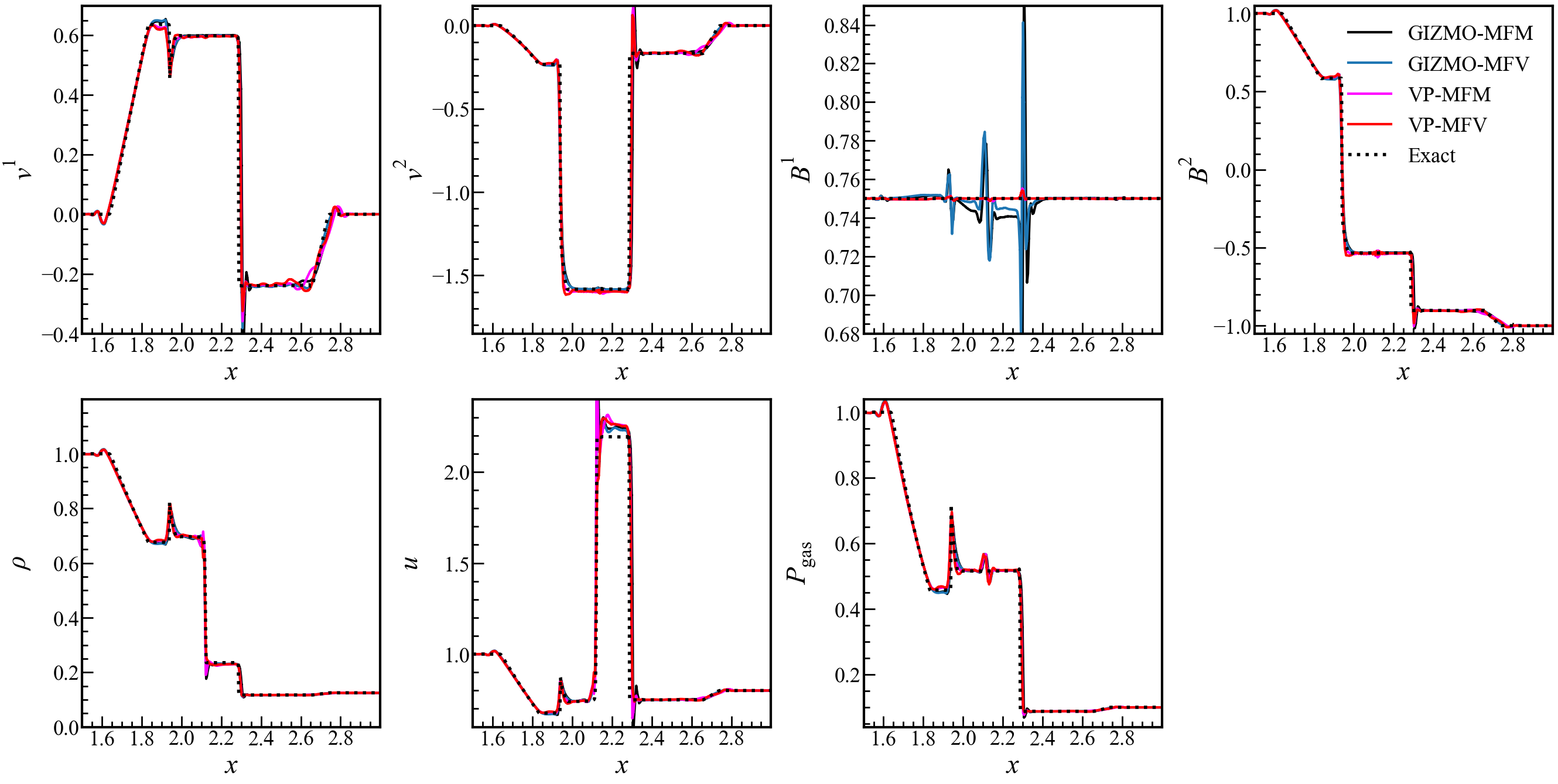}
\caption{{\small Brio-Wu shock tube. Density, velocity, magnetic field, internal energy and pressure is evolving by GIZMO-MFM method (dark line), GIZMO-MFV method (blue line), VP-MFM method (purple line), VP-MFV method (red line)  and exact solution (dashed line)  at $t=0.2$, respectively.}}
\label{Fig:1}
\end{figure*}

This test is designed to measure the accuracy of MHD shocks, rarefaction wave and contact discontinuities. Compared with GIZMO-MFM and GIZMO-MFV methods, the results of density, energy, pressure and velocity evolution of the VP-MFM and VP-MFV methods are consistent. There still exist oscillation at boundaries of the shock and discontinuities, but the positions are accurately captured, the magnitude of oscillations are similar with the GIZMO results. It is obvious that a significant improvement on the $B^1$ occurs, when comparing to the GIZMO in Fig.~\ref{Fig:1}. Note, the magnetic oscillation of the VP method at $x \approx 2.1$ is caused by the non-smooth initial vector potential.

\subsection{2D Orszag-Tang vortex}
\label{sec:2dotv}
The second test is the 2-D compressible Orszag-Tang vortex problem\cite{Orszag+1979}. In a periodic squared domain, we set initial parameters to $\gamma = 5/3$, $\rho = {25}/{36\pi}$, $P_{gas} = {5}/{12\pi}$, $(v^1,v^2) = \{ -\sin(2\pi y), \sin(2\pi x) \}$, and $(B^1,B^2) = \{ -\sin(2\pi y)/\sqrt{4\pi}, \sin(4\pi x)/\sqrt{4\pi} \}$. Correspondingly, the vector potential is initialized to $A^3 = 1/(8\pi^{3/2})[2\cos(2\pi y) + \cos(4\pi x)]$. All simulations contain $256^2$ particles.

\begin{figure*}[htpb] 
\centering
\includegraphics[width=0.49\linewidth]{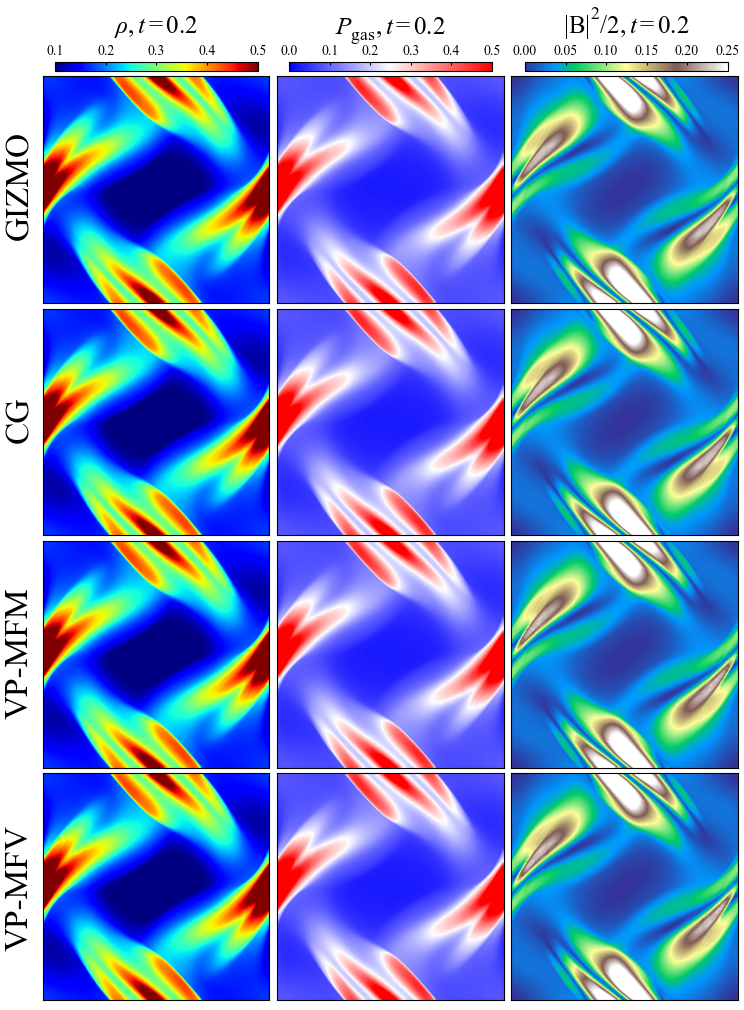}
\includegraphics[width=0.49\linewidth]{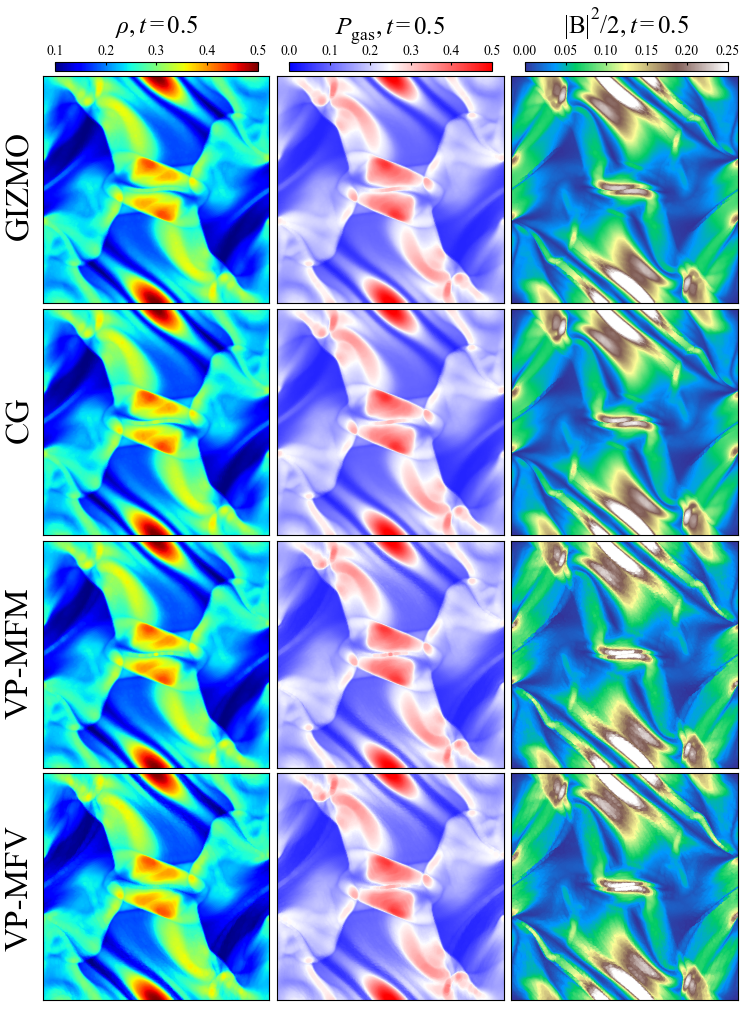}
\caption{{{The 2D Orszag-Tang vortex. The distribution of $\rho$, $P_{gas}$ and $|\mathbf{B}|^2/2$ shows in the columns from left to right, respectively. For comparison, the GIZMO case is presented in the first row, the CG is in the second, the VP-MFM is in the third and VP-MFV is in the fourth row. }}}
\label{Fig:2}
\end{figure*}

In Fig.~\ref{Fig:2}, we present the distribution of density, pressure and magnetic fields of simulations run with GIZMO, {CG, VP-MFM and VP-MFV} at $t = 0.2, 0.5$. Both VP-MFM and VP-MFV are highly consistent with the GIZMO {and CG} results. 

\begin{figure*}
\centering
\includegraphics[width=0.49\linewidth]{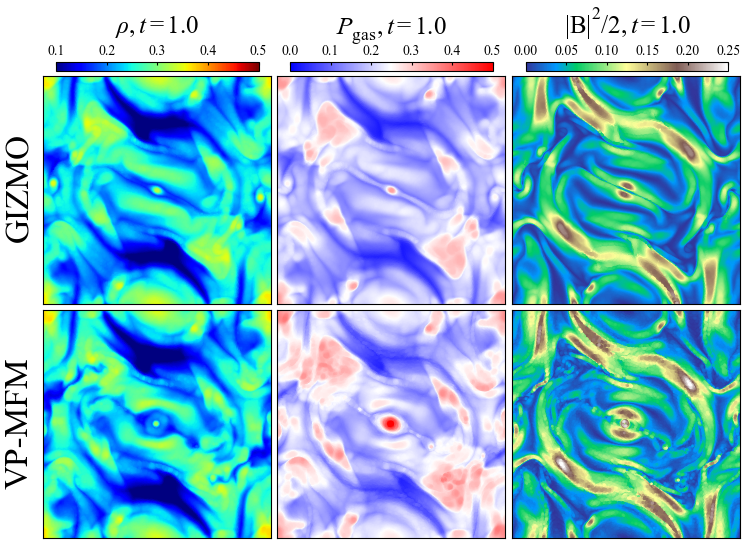}
\includegraphics[width=0.49\linewidth]{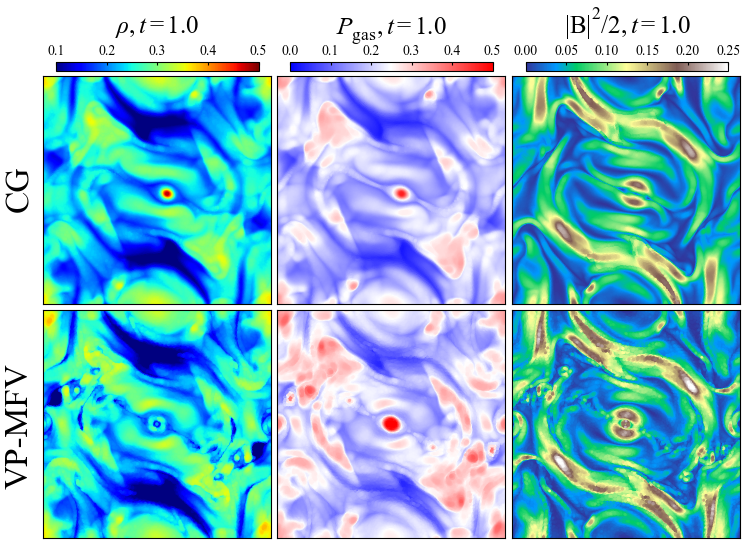}
\caption{{The 2D Orszag-Tang vortex. Distribution of $\rho$, $P_{gas}$ and $|\mathbf{B}|^2/2$ of 2D Orszag-Tang vortex at time $t=1.0$ in GIZMO, CG, VP-MFM and VP-MFV methods. } }
\label{Fig:3}
\end{figure*}

However, the deviation of VP and GIZMO method becomes visible at time $t = 1$. In Fig.~\ref{Fig:3}, we present distribution of mass density, gas pressure and magnetic energy density of simulations run with GIZMO, GIZMO with CG and VP in MFM and MFV method, respectively. All three methods can resolve the central vortex (density and pressure), but the vortex size and strength are quite different. Especially in terms of pressure $P_{gas}$, while the central vortex pressure $P_{gas}$ of VP-MFM and VP-MFV are close to 0.68 which is {close to} the results of mesh code~\citep{Stone2008} (e.g., by the Athena with Roe solver and the third order algorithm\footnote{\url{https://www.astro.princeton.edu/~jstone/Athena/tests/orszag-tang/pagesource.html}}), the feature of central vortex in GIZMO is less prominent than our method. In GIZMO-CG method, while the 
amplitude of magnetic field is improved in relative to GIZMO-MFM, the vortex offset the centre moderately. In addition, more small vortexes near boundaries are observed in VP method than GIZMO or GIZMO-CG.

\begin{figure}
\centering
\includegraphics[width=0.8\linewidth]{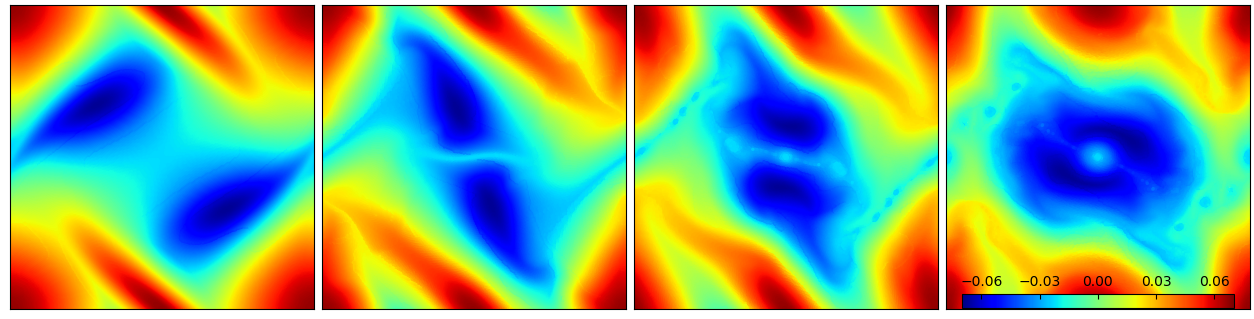}
\caption{Distribution of the vector potential $A^3$ in 2D Orszag-Tang vortex case is evolved by VP-MFV at time $t= 0.2, 0.5, 0.7$ and $1.0$, from left to right panels. }
\label{Fig:4}
\end{figure}

\begin{figure}[htpb] 
\centering
\includegraphics[width=0.98\linewidth]{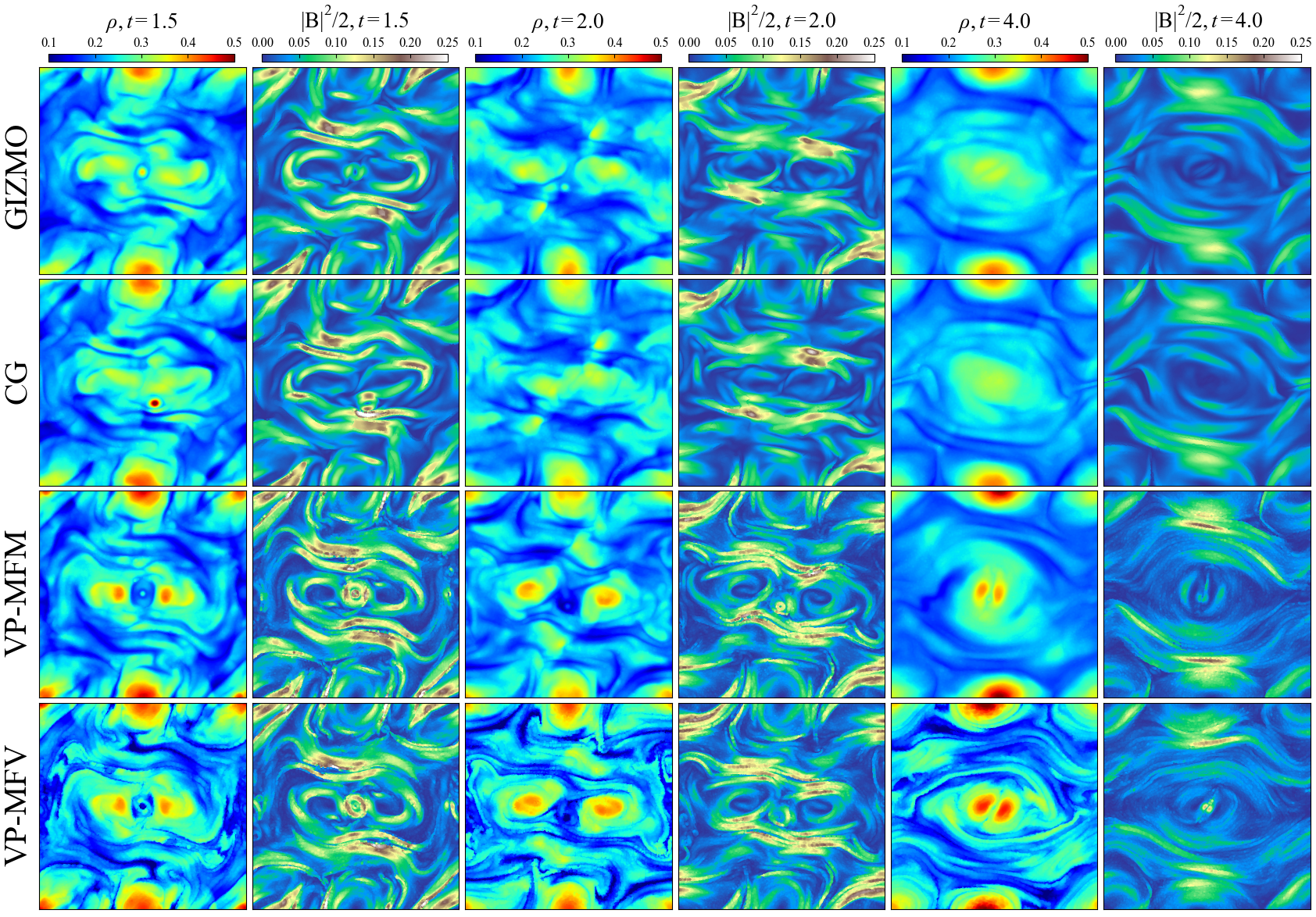}
\caption{{ The 2D Orszag-Tang vortex. Distribution of $\rho$ and $|\mathbf{B}|^2/2$ of 2D Orszag-Tang vortex, $t= 1.5, 2.0$ and $4.0$ from left to right panels, in GIZMO, CG, VP-MFM and VP-MFV methods.  }}
\label{Fig:5}
\end{figure}

It is not surprising that the VP method produces such better results. As described in Section~\ref{sect:evp},  in our method, the magnetic field $\mathbf{B}$ is exactly zero-divergence and has analytical solutions in 2-D cases, therefore stable long-term evolution of magnetic field $\mathbf B$ is expected. Fig.~\ref{Fig:4} displays the evolution of the vector   potential $A^3$ based on VP-MFV method at time $t= 0.2, 0.5, 0.7, 1.0$. It is obvious that the symmetry and details at small scales can be precisely captured.

In Fig.~\ref{Fig:5}, we present the distribution of density and magnetic pressure $|\mathbf{B}|^2/2$ of simulations run with GIZMO, CG, VP-MFM and VP-MFV at $t = 1.5, 2.0, 4.0$, respectively. The result of CG quickly break the central symmetry. The tiny residual of magnetic divergence in CG could cause this long term deviation. The density magnitude and $|\mathbf{B}|^2/2$ of VP keep much larger than CG and GIZMO. 

\subsection{3D Orszag-Tang vortex}

The test problem we conduct  {in this subsection} is the 3D Orszag \& Tang vortex. We follow \citet{Helzel+2011} to set up numerical parameters,  namely in a periodic 3D unit box, the initial condition is set to be $\rho = {25}/{36\pi}$, $P_{gas} ={5}/{12\pi}$, 
\begin{equation}\notag
(v^1,v^2,v^3) = \{-(1+\epsilon_p\sin(2\pi z))\sin(2\pi y),  (1+\epsilon_p\sin(2\pi z))\sin(2\pi x), \epsilon_p\sin(2\pi z) \}
\end{equation}
\begin{equation}\notag
(B^1,B^2,B^3) = \{ -\frac{\sin(2\pi y)}{\sqrt{4\pi}}, \frac{\sin(4\pi x)}{\sqrt{4\pi}}, 0 \},
\end{equation}
and the initial magnetic field derived by $A^3 = 1/(8\pi^{3/2})[2\cos(2\pi y) + \cos(4\pi x)]$ {in the vector potential formulation}. Here we choose $\epsilon_{p} = 0.2$. The adiabatic index of gas is $\gamma = 5/3$. We perform 4 simulations from the same initial setup but with different MHD schemes. All runs are carried out with $128^3$ particles.

\begin{figure*}[htpb] 
\centering
\includegraphics[width=0.95\linewidth]{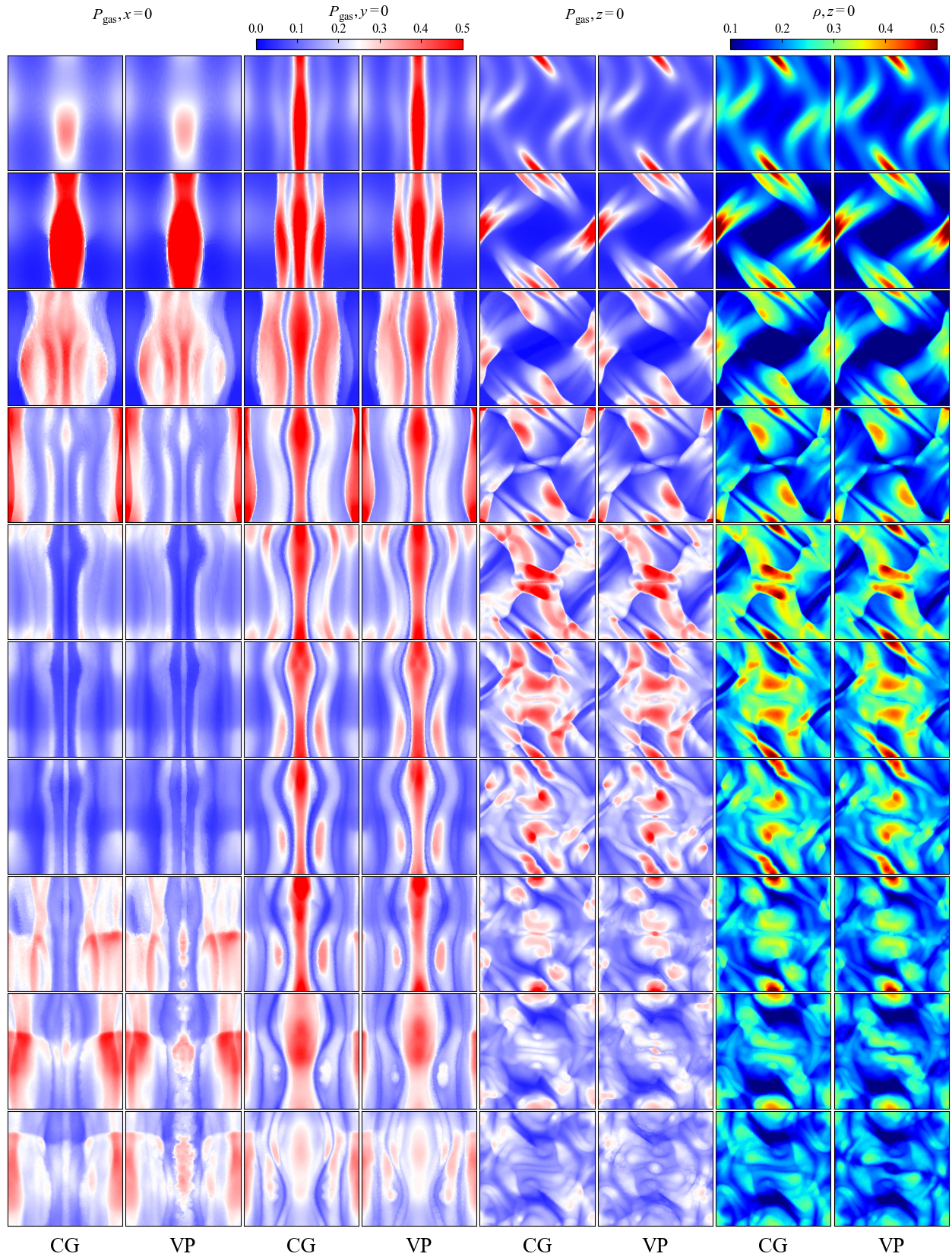}
\caption{Comparison of the pressure $P_{gas}$ and density $\rho$ between CG (column 1,3, 5 and 7) and VP (column 2, 4, 6 and 8) at time from $t = 0.1$ (top) to 1 (bottom row).}
\label{Fig:6}
\end{figure*}

Fig.~\ref{Fig:6} shows the evolution till $t=1$. Due to the periodic condition, we list the pressure at three surfaces $x=0, y=0, z=0$ and density at the surface of $z=0$ for comparison. The results of the VP method are in good agreement with the results of the CG method before $t=0.5$. After that, the difference between VP method and CG method increases gradually. At time $t=1$, a clear vortex appears in VP method, but failed in CG. 

\begin{figure}[htpb] 
\centering
\includegraphics[width=\linewidth]{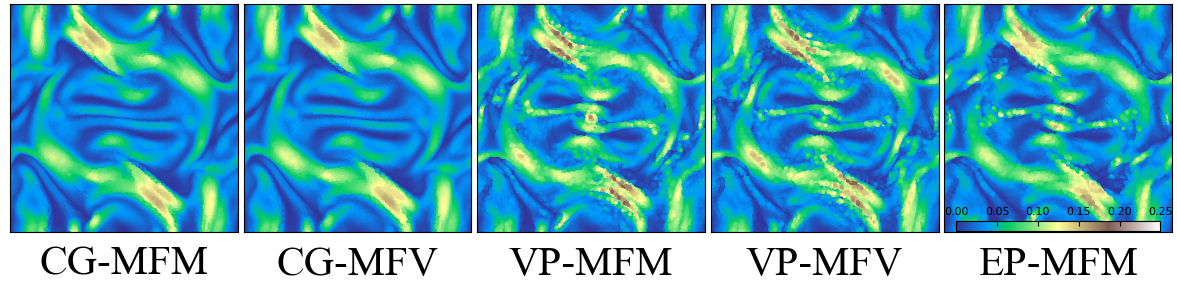}
\caption{{Evolution of 3D Orszag-Tang vortex. The panels from left to right show the magnetic pressure $|\mathbf{B}|^2/2$  in the side of $z=0$ at $t=1.0$ in  CG-MFM, CG-MFV, VP-MFM, VP-MFV method and EP-MFM, respectively. }}
\label{Fig:7}
\end{figure}

\begin{figure}[htpb] 
\centering
\includegraphics[width=0.6\linewidth]{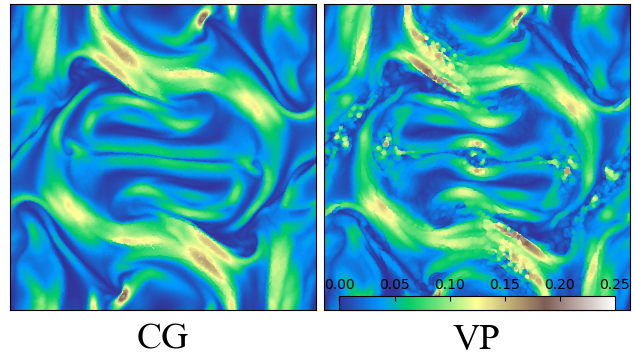}
\caption{{Evolution of 3D Orszag-Tang vortex. The panels show the magnetic pressure $|\mathbf{B}|^2/2$  in the side of $z=0$ at $t=1.0$ with $256^3$ resolution in CG and VP methods, respectively. }}
\label{Fig:8}
\end{figure}

\begin{figure}[htpb] 
\centering
\includegraphics[width=\linewidth]{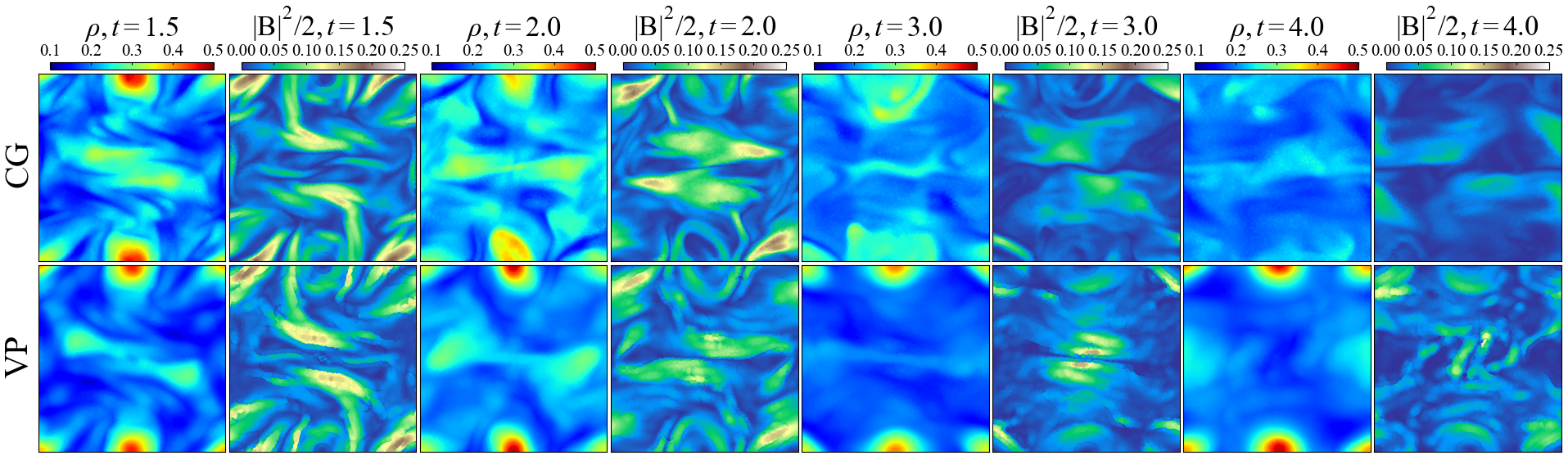}
\caption{{Evolution of 3D Orszag-Tang vortex. The panels show $\rho$ and the magnetic pressure $|\mathbf{B}|^2/2$  in the side of $z=0$ at $t=1.5,2.0,3.0$ and $4.0$ from left to right panels, in CG and VP methods, respectively. }}
\label{Fig:9}
\end{figure}

In Fig.~\ref{Fig:7}, we present  the vortex structure of magnetic pressure with different methods at $t=1$. The VP methods give rise to small scale structures in the central regions, which are not present in the CG methods. Note that there is also small difference between the VP-MFM and VP-MFV.

In this specific case, we employed another approach, Euler potentials (EP \citep{Rosswog2007,Brandenburg2010,Price+2007}), to check the central structure. Besides the vector potential, magnetic field can be written as $\mathrm{B}=\nabla \alpha \times \nabla \beta$, where the $\alpha$ and $\beta$ are scalar fields called Euler potentials. For the idea magnetodynamics with zero resistivity, the EP hold simple dynamic equations,
\begin{equation}
\dfrac{d \mathbf{\alpha}}{d t} = 0, ~~~ \dfrac{d \mathbf{\beta}}{d t} = 0.
\end{equation}
In our Lagrangian approach, they are numerically discreted to $\alpha_i^{n+1} = \alpha_i^{n}$ and $\beta_i^{n+1} = \beta_i^{n}$ and the initial values are $\alpha^0=A_0^3$ and $\beta^0=z_0$. This exact results replace the dynamic equation  (Eq.~\ref{eq:discrete2}) to compute the discrete magnetic field $B_d$. The result is presented in the last panel of Fig.~\ref{Fig:7}. Since EP are the simplest and exact, the EP result is trusty. Obviously, the result of EP-MFM is more consistent with our VP methods than CG-MFM or CG-MFV.

Further, we carried out a $256^3$ implementation. In Fig.~\ref{Fig:8}, the higher resolution result obtains a clearer solutions. It supports some details of vortex in $128^3$ run are not noise.

{In order to estimate the long term stability, we carried out the 3D VP-MFM test to 4 time units, comparing with CG-MFM.  In Fig.~\ref{Fig:9}, the density $\rho$ and the magnetic pressure $|\mathbf{B}|^2/2$ at the slice of $z=0$  are present till $t=4$ in CG and VP methods, at top panels and bottom panels, respectively. After $t=2$, the difference between two schemes become clear and the shock structure and amplitude of density and magnetic fields are larger for VP run. This test demonstrates that VP scheme can keep numerically stable even for a very long-term evolution.
}

\subsection{Magnetorotational Instability}
In this section, we examine the VP scheme for a crucial numerical experiment in  astrophysics \citep{Balbus+1991,Balbus+1998,Guan+2008}, Magnetorotational Instability (MRI). MRI is carried out in  a axisymmetric 2D shearing box, which is a model of the $x-z$ cross section of the magnetized disks with angular speed of $\Omega=1$. In an axisymmetric squared domain, $-0.5<x<0.5$ and $-0.5<z<0.5$. We setup the initial condition of density $\rho=1$, pressure $P_{gas}=1$, velocity $(v^1,v^2,v^3)=\{\delta v,-q x, \delta v\}$ and magnetic field $(B^1,B^2,B^3)=\{0,0,B_{0}\sin(2\pi x)\}$, where $ q =3/2, \gamma = 5/3$ and  $B_0=\sqrt{15}/(8\pi m)$ with the mode number $m=4$, following \citet{Hopkins+2016a}. The velocity fluctuation is randomly generated within the range of  $\delta v \in[-0.005,0.005]$ . Correspondingly, the vector potential in our scheme could be initialized as $A^2=-B_{0}\cos(2\pi x)/(2\pi)$. In this test, we have to consider the momentum equations with an additional source terms, $D(\rho \mathbf{v})/Dt = -2(\Omega\hat{z})\times(\rho \mathbf{v}) + 2\rho q \Omega^2x\hat{x}$, which corresponds to the Coriolis and the centrifugal forces, respectively. All variables employ periodic boundary in the direction of $x$ and $z$, except for $v^2$, $A^1$ and $A^3$ in $x$ axis. In specific, $v^2(x,z)=v^2(x+ n_x L_x,z+ n_z L_z) + n_x q \Omega L_x$, $A^1(x,z) = -A^1(x+ n_x L_x,z+ n_z L_z)$, and  $A^3(x,z) = -A^3(x+ n_x L_x,z+ n_z L_z)$. 

\begin{figure}[htpb] 
\centering
\includegraphics[width=0.8\linewidth]{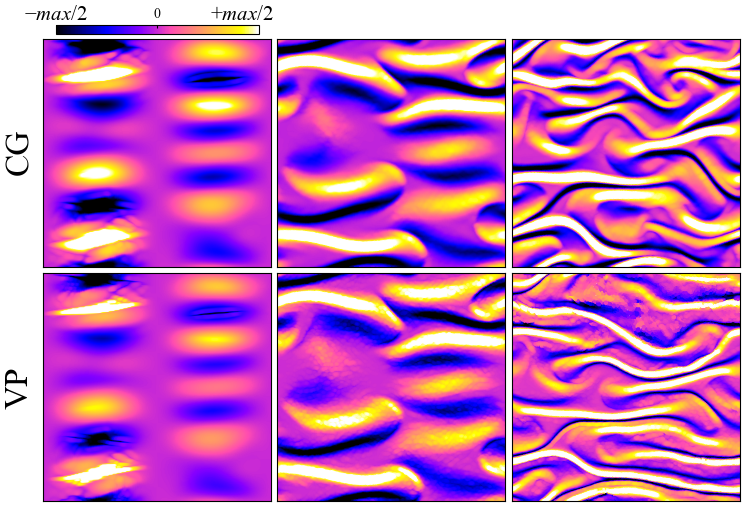}
\caption{{Evolution of Magnetorotational Instability. The panels show the component $B^1$ of the magnetic field at time $t= 10, 16$ and $18$ from left to right panels in CG and VP methods, respectively. }}
\label{Fig:10}
\end{figure}

\begin{figure}[htpb] 
\centering
\includegraphics[width=0.8\linewidth]{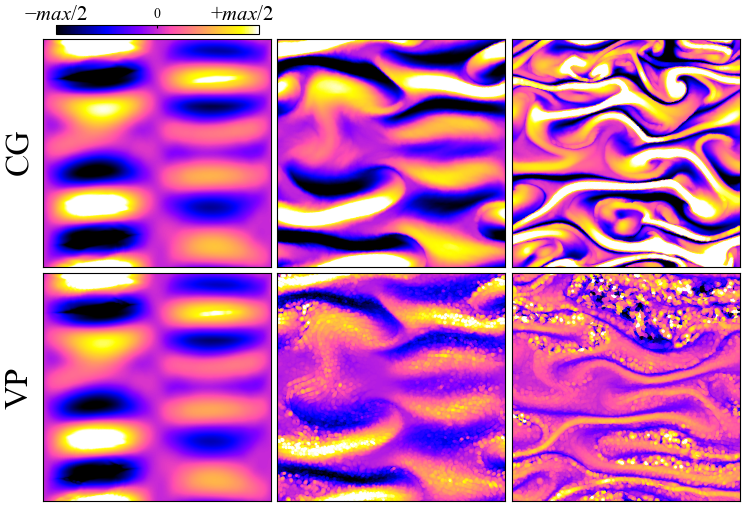}
\caption{{Evolution of Magnetorotational Instability. The panels show the component $B^2$ of the magnetic field at time $t= 10, 16$ and $18$ from left to right panels in CG and VP methods, respectively. }}
\label{Fig:11}
\end{figure}

The evolution of magnetic field $B^1$ and $B^2$ are presented in Fig.~\ref{Fig:10} and Fig.~\ref{Fig:11}, respectively. There are clear $m=4$ pattern at $t=10$ and systems are transforming into turbulent mode at $t \ge 16$. The pattern of VP is highly consistent with the CG scheme. Note that the behavior of $B^2$ is more noisy than $B^1$. We confirm the source of noise is due to the discontinuity of VP at $x$-axes boundary. In CG approach, the dynamic variable is magnetic field itself and $B^1$, $B^2$ and $B^3$ are artificially satisfied with the periodic condition. But it is not periodic for the vector potentials $A^1$ and $A^3$ with the VP scheme in this test. We have to allow a natural extension of VP at the outer region of $x$-axes boundary. However this extension will cause some noise on the numerical derivative. Despite we cannot find out a better solution to deal with this periodic boundary, we deduce that VP method will perform well under a physical boundary.

\section{Conclusion}
\label{sect:conclusion}
We developed a novel meshless method based on vector potential of magnetic field in MHD. In this method, the vector potential ${\mathrm A}$ is used as a primitive variable, instead of magnetic field ${\mathrm B}$, and a conservative Riemann solver is adopted for flux calculation. Such a scheme automatically vanishes the divergence and does not need the extra artificial correction, while the derived magnetic field need to be stabilized by smoothing. 

We test our scheme by running the Brio-Wu shock tube, two and three dimensional Orszag-Tang vortex test problems {and Magnetorotational Instability}, and compare our results with those of the GIZMO implementations. In the Brio-Wu test, the numerical oscillation of magnetic field at discontinuous point in GIZMO almost disappears in our method. In the 2D vortex test, the magnetic field in our method perfectly forms a central vortex and captures detailed pattern in the entire computing zone, much better than the results of GIZMO. In the 3D vortex experiment, both amplitude and position of vortex are more clear, symmetric with our method than their counterparts with CG-GIZMO. The CG method calibrates the flux of magnetic field over compact boundaries, which inevitably distorts the derivative of $\mathbf{B}$. These errors do not show up in our method. On the other hand, our method employs a Godunov scheme with advantage of treating disconnection and keeping conservation when comparing with a VP-SPH method.

Finally, we conclude that our method is effective and robust, and it provides a better control on magnetic field divergence, especially on the discontinuous points and shock wave front. Meanwhile, VP-MFM/MFV is flexible to be implemented for programming and it is potentially powerful for high dynamic range MHD simulation.

\section{Acknowledgements}
We acknowledge the support from National SKA Program of China (Grant No. 2020SKA0110401), Special Research Assistant Program of the Chinese Academy of Sciences and K.C.Wong Education Foundation. LG acknowledge support from NSFC grant (No. 11425312), and two Royal Society Newton Advanced Fellowships, as well as the hospitality of the Institute for Computational Cosmology at Durham University.

\end{document}